\documentclass[11pt]{article}
\usepackage{blois,epsfig}

\usepackage{amsmath}
\newcommand{\dphi}{\ensuremath{\Delta\varphi}}
\newcommand{\pTassoc}{\ensuremath{p_{\mathrm{T,\,assoc}}}}
\newcommand{\pTtrig}{\ensuremath{p_{\mathrm{T,\,trig}}}}
\newcommand{\sqrts}{\ensuremath{\sqrt{s}}}
\newcommand{\Nch}{\ensuremath{N_{\mathrm{ch}}}}
\newcommand{\Nnear}{\ensuremath{\langle N_{\mathrm{assoc,\, near\ side}} \rangle}}
\newcommand{\Naway}{\ensuremath{\langle N_{\mathrm{assoc,\, away\ side}} \rangle}}
\newcommand{\Nisotrop}{\ensuremath{\langle N_{\mathrm{isotrop}} \rangle}}
\newcommand{\Ntrigger}{\ensuremath{\langle N_{\mathrm{trigger}} \rangle}}
\newcommand{\NuncorrAv}{\ensuremath{\langle N_{\mathrm{uncorrelated\, seeds}} \rangle}}
\begin{document}
\vspace*{4cm}
\title{MULTIPLICITY DEPENDENCE OF TWO-PARTICLE CORRELATIONS IN PROTON-PROTON COLLISIONS MEASURED WITH ALICE AT THE LHC}

\author{ Eva Sicking on behalf of the ALICE Collaboration }

\address{European Organization for Nuclear Research (CERN), Geneva, Switzerland and \newline 
Institut f\"ur Kernphysik, Westf\"alische Wilhelms-Universit\"at M\"unster, M\"unster, Germany \newline
eva.sicking@cern.ch}

\maketitle
\begin{figure}
\centering
\psfig{figure=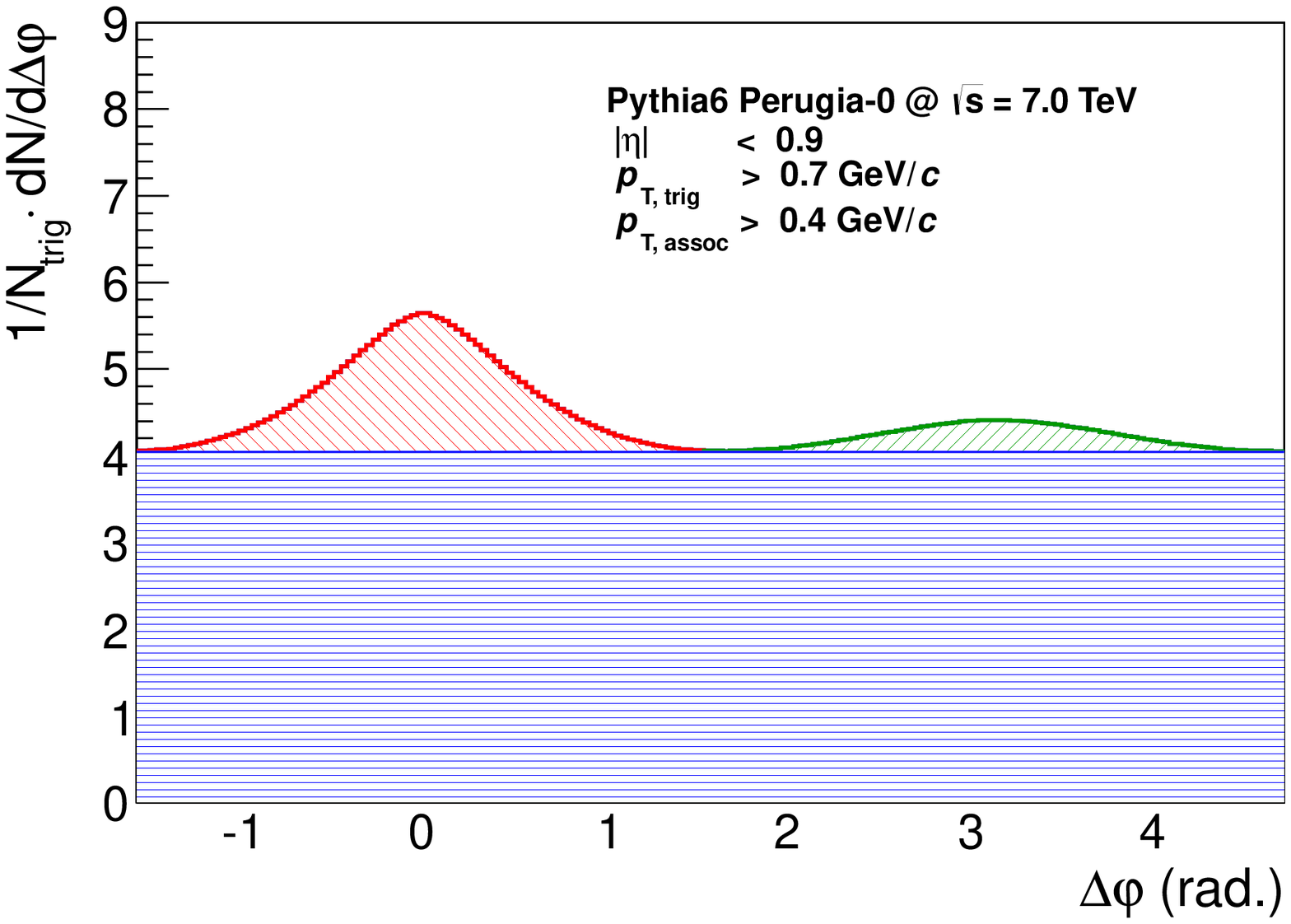,width=75mm}
\psfig{figure=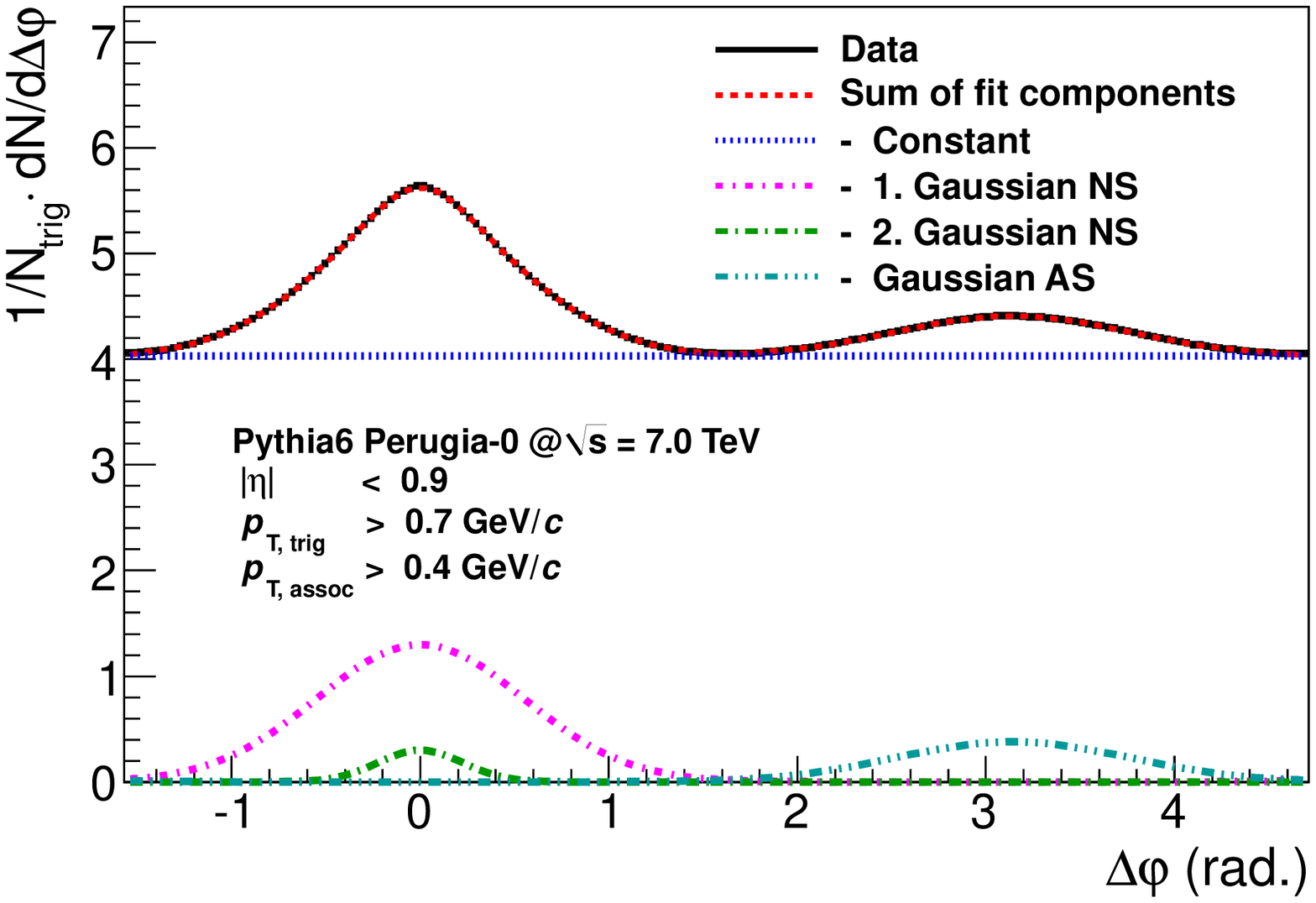,width=75mm}
\caption{Left panel: Geometrical contributions to the per-trigger yield as a function of $\dphi$.
Right panel: The per-trigger yield as a function of $\dphi$ described by a fit function and its sub-components. }
\label{fig:dphiExtract}
\end{figure}

\abstracts{
We investigate properties of jets in proton-proton collisions using two-particle angular correlations. By choosing an analysis approach based on two-particle angular correlations, also the properties of low-energetic jets can be accessed. Observing the strength of the correlation as a function of the charged particle multiplicity reveals jet fragmentation properties as well as the contribution of jets to the overall charged particle multiplicity. Furthermore, the analysis discloses information on the underlying multiple parton interactions. We present results from proton-proton collisions at the center-of-mass energies $\sqrt{s} = 0.9$, 2.76, and 7.0\,TeV recorded by the ALICE experiment. The ALICE data are compared to Pythia6, Pythia8, and Phojet simulations.
}
We study two-particle angular correlations in azimuthal direction induced by low energetic hadronic di-jets produced in proton--proton (pp) collisions.
Using two-particle angular correlations with low transverse momentum thresholds, it is possible to measure jet properties on a statistical basis down to the lowest possible jet energies. In contrast, the reconstruction of jets on an event-by-event basis only allows to access jets of energies significantly above the underlying event contribution \cite{ALICE:2011ac} that can be large at high particle multiplicities. In addition, two-particle correlations are insensitive to the overlap of jets which has a sizable probability at high multiplicities \cite{Abelev:2012ej}. The presented results allow to optimize the parametrization of the jet fragmentation in Monte Carlo generators down to low energies into the non-perturbative regime and can give information about the contribution from multiple parton interactions.\\
We are plotting the azimuthal difference between trigger particles (defined by $\pTtrig>0.7$\,GeV/$c$, $\arrowvert \eta \arrowvert<0.9$) and associated particles (defined by $\pTassoc>0.4$\,GeV/$c$, $\arrowvert \eta \arrowvert<0.9$). The main observable of the two-particle correlation is the per-trigger yield as a function of the difference in the azimuthal~angle
\begin{equation}
\frac{\mathrm{d}N}{\mathrm{d}\dphi} = \frac{1}{N_{\mathrm{trig}}} \frac{\mathrm{d}N_{\mathrm{assoc}}}{\mathrm{d}\dphi},
\end{equation}
where $N_{\mathrm{trig}}$ is the number of trigger particles, $\dphi = \varphi_{\mathrm{trig}}-\varphi_{\mathrm{assoc}}$ is the difference in the azimuthal angle of the two particles, and $N_{\mathrm{assoc}}$ is the number of associated particles. The analysis allows to study the per-trigger yield as a function of the charged particle multiplicity as well as for different transverse momentum thresholds $\pTtrig$ and $\pTassoc$. This results in a vast amount of different collision event classes and particle-pair classes.\\
In order to extract the per-trigger yields for all classes, a fit function is introduced which allows to decompose the azimuthal correlation into its main components. The left panel of Figure~\ref{fig:dphiExtract} shows that the per-trigger yield as a function of $\dphi$ can be divided into a flat combinatorial background of the azimuthal correlation, a peak at the near side of the trigger particles ($\dphi \approx0$\,rad), and a peak at the away side of the trigger particles ($\dphi \approx\pi$\,rad).

\begin{figure}[h]
\begin{center}
\psfig{figure=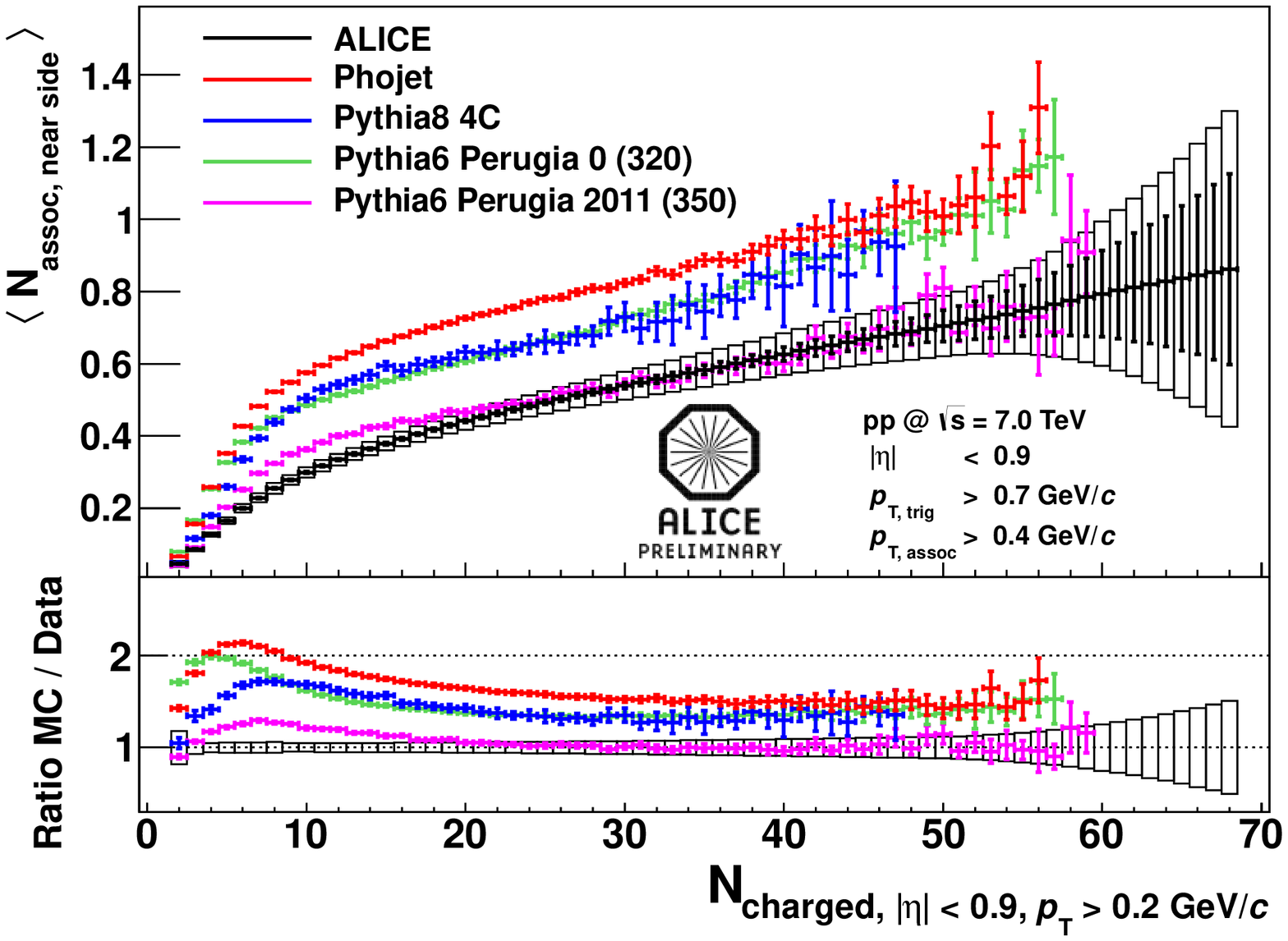,width=75mm}
\psfig{figure=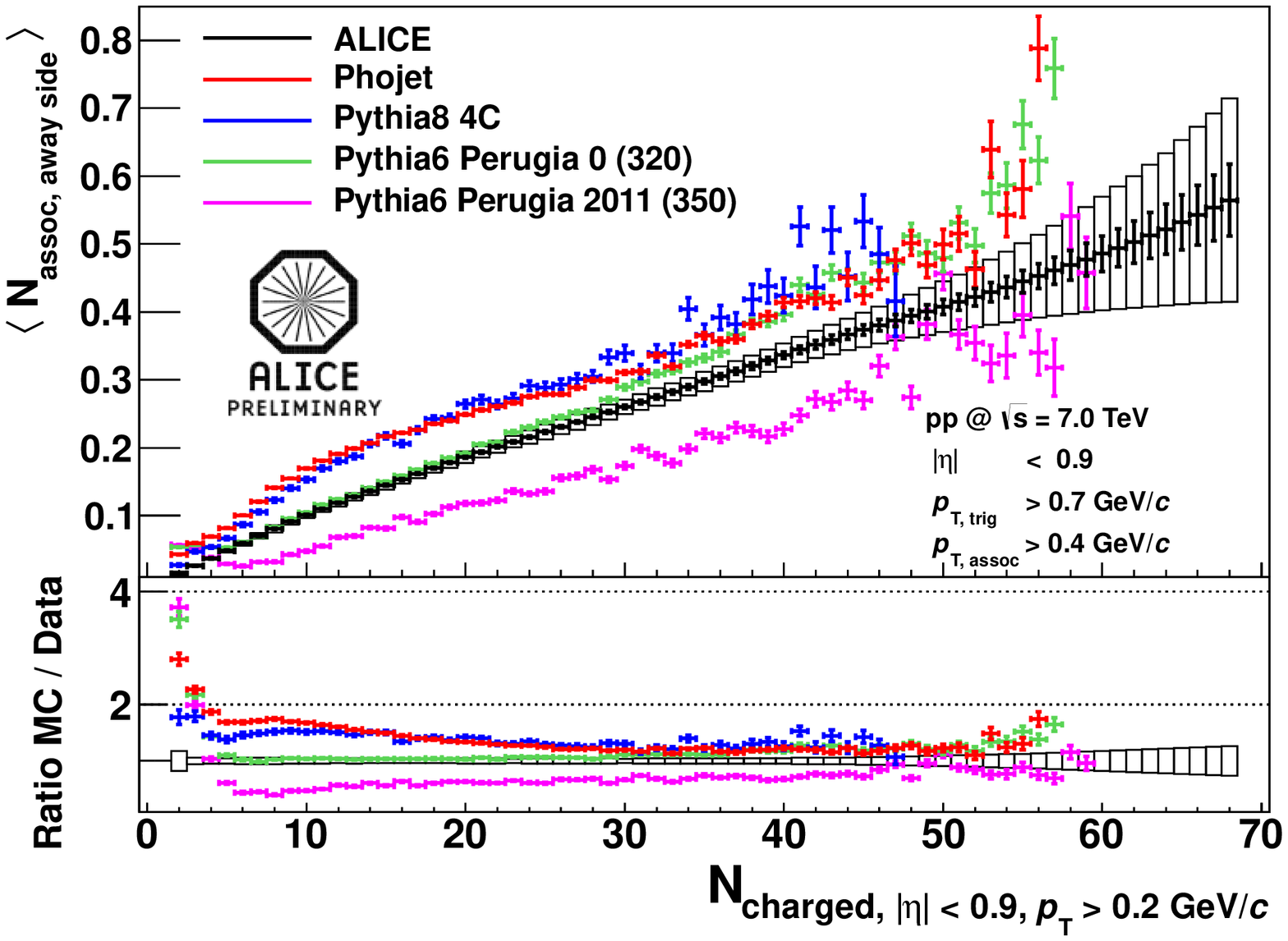,width=75mm}
\caption{Integrated per-trigger yield in the near side peak (left panel) and in the away side peak (right panel) above the combinatorial background as a function of the charged particle multiplicity measured at $\sqrts=7.0$\,TeV. The error bars of the data points represent the statistical uncertainties. The systematic uncertainties are presented as boxes around the ALICE data points.}
\label{fig:near_away}
\end{center}
\end{figure}

\begin{figure}[h]
\begin{center}
\psfig{figure=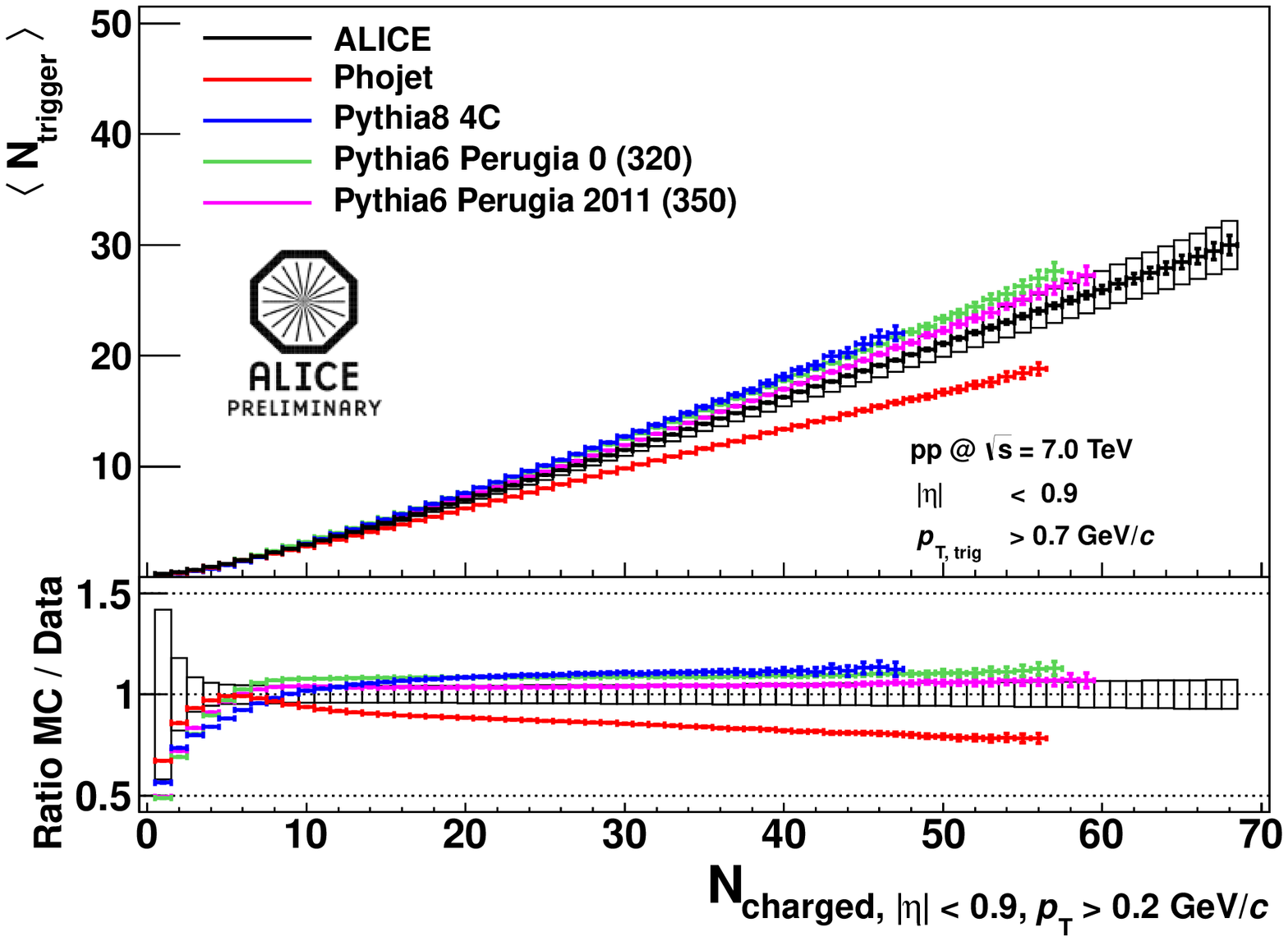,width=75mm}
\psfig{figure=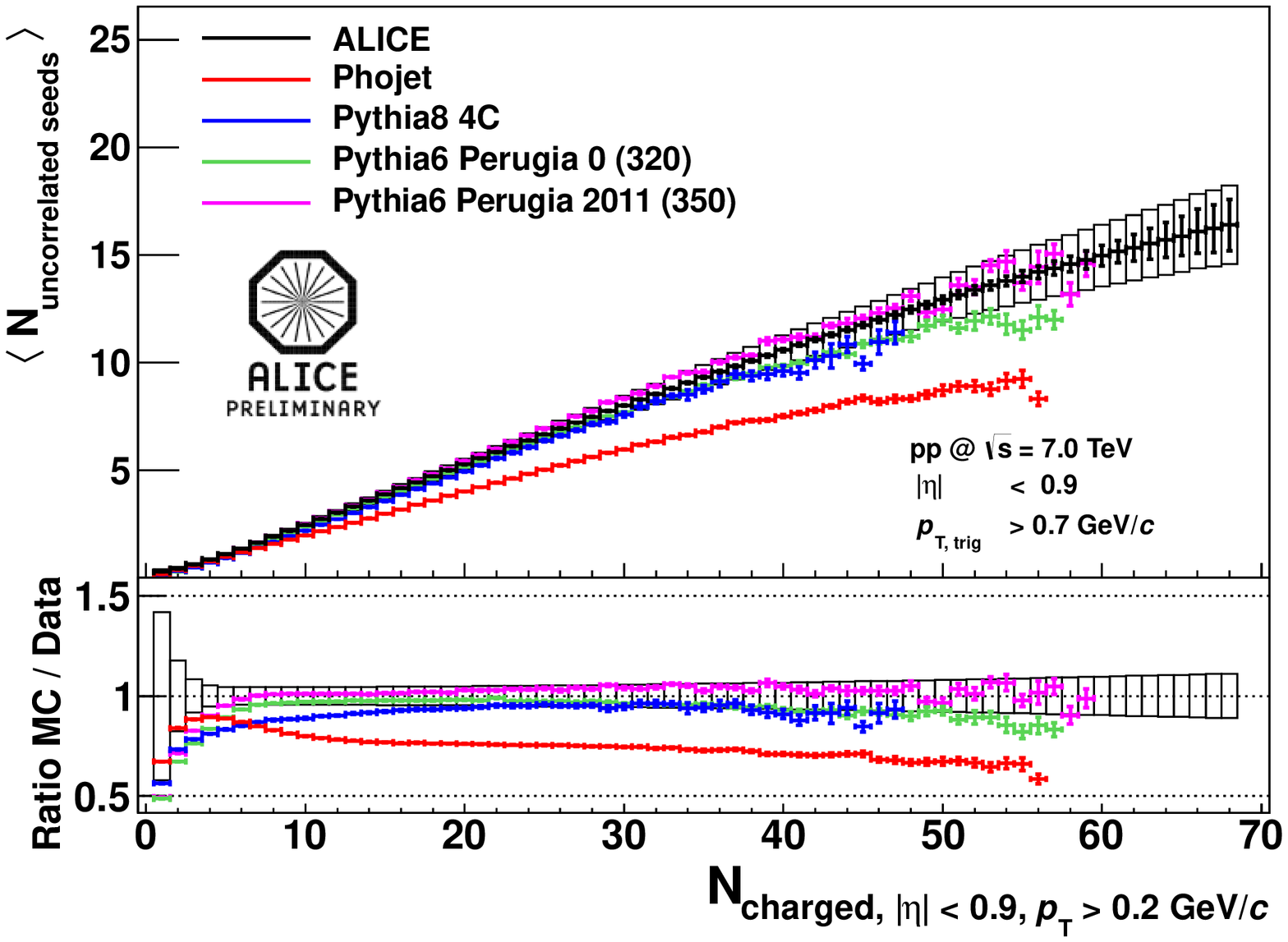,width=75mm}
\caption{Average number of trigger particles (left panel) and average number of uncorrelated seeds (right panel) as a function of the charged particle multiplicity measured at $\sqrts=7.0$\,TeV.}
\label{fig:trigger_nuncorr}
\end{center}
\end{figure}

The azimuthal correlation can very well be approximated by a fit function comprising a constant for the combinatorial background, two Gaussian functions for the near side peak, and one Gaussian function for the away side peak. The right panel of Figure~\ref{fig:dphiExtract} shows the measured azimuthal correlation, the parametrization of the correlation based on the fit function, and the sub-components of the fit function. \\
Based on the fit function, five observables can be derived from the azimuthal correlation. Three of the observables represent the decomposed per-trigger yield: the per-trigger yield in the combinatorial background $\Nisotrop$, the per-trigger yield in the near side peak $\Nnear$, and the per-trigger yield in the away side peak $\Naway$. In addition, the average number of trigger particles $\Ntrigger$ is measured. These observables are combined to the average number of uncorrelated seeds $\NuncorrAv$, a measure of the number of uncorrelated sources of particle production. $\NuncorrAv$ is defined as
\begin{eqnarray}\label{eq:uncorr}
\NuncorrAv = \frac{\langle N_{\mathrm{trigger}} \rangle}{\langle 1 + N_{\mathrm{assoc, \; near+away}, \;p_{\mathrm{T}}>p_{\mathrm{T,\; trig}}}\rangle}.
\end{eqnarray}
It can be demonstrated in Pythia simulations that the average number of uncorrelated seeds is proportional to the average number of multiple parton interactions (figure not shown).\\
The analysis is used to analyze minimum bias pp collision data recorded by the ALICE (A Large Ion Collider Experiment) experiment~\cite{Aamodt:2008zz}. ALICE has recorded pp collision data sets at three different collision energies, $\sqrts=0.9$, 2.76, and 7.0\,TeV. For the vertex finding and the tracking, the two central tracking detectors of ALICE are used, the Inner Tracking System~(ITS) and the Time Projection Chamber~(TPC). The collision data have been corrected for detector effects.
The ALICE results are compared to simulations of Pythia6.4 \cite{Sjostrand:2006za} (tune Perugia-0 \cite{Skands:2010ak} and tune Perugia-2011 \cite{Skands:2010ak}), Pythia8.1 \cite{Sjostrand:2007gs} (tune 4C), and Phojet~\cite{Engel:1995sb} (version~1.12).\\
Due to the limited space, in the following only few significant figures estimated at $\sqrts=7.0$\,TeV, $\pTtrig>0.7$~\,GeV/$c$, and \mbox{$\pTassoc>0.4$~\,GeV/$c$} are discussed exemplarily.\\
The per-trigger yield in the combinatorial background of the correlation grows as a function of the charged particle multiplicity. The ALICE results are well described by all models within the uncertainties for all charged particle multiplicities (figure not shown).\\
The per-trigger near side yield dominated by the fragmentation of jets is presented in the left panel of Figure~\ref{fig:near_away}. At low multiplicities, all models overestimate the near side yield by 40\,\% to 120\,\%. Pythia6 Perugia-2011 gives the best description of the ALICE data. The agreement between the models and the ALICE data improves with increasing charged particle multiplicity.\\
The per-trigger away side yield dominated by the fragmentation of recoiling jets is presented in the right panel of Figure~\ref{fig:near_away}. The difference between the models and the ALICE results varies between -50\,\% and +100\,\%. Only Pythia6 Perugia-0 agrees within the uncertainties with the ALICE results for charged particle multiplicities below $\Nch=30$.\\
The average number of trigger particles as a function of the charged particle multiplicity is presented in the left panel of Figure~\ref{fig:trigger_nuncorr}. All Pythia tunes slightly overestimate the ALICE results and Phojet underestimates the ALICE results. The average number of uncorrelated seeds (c.f.\ Equation~\ref{eq:uncorr}) is presented in the right panel of Figure~\ref{fig:trigger_nuncorr}. Whereas the ALICE results are significantly underestimated by Phojet, Pythia reproduces the results with good agreement.

\vspace{-0.6cm}

\begin{figure}[h]
\begin{center}
\psfig{figure=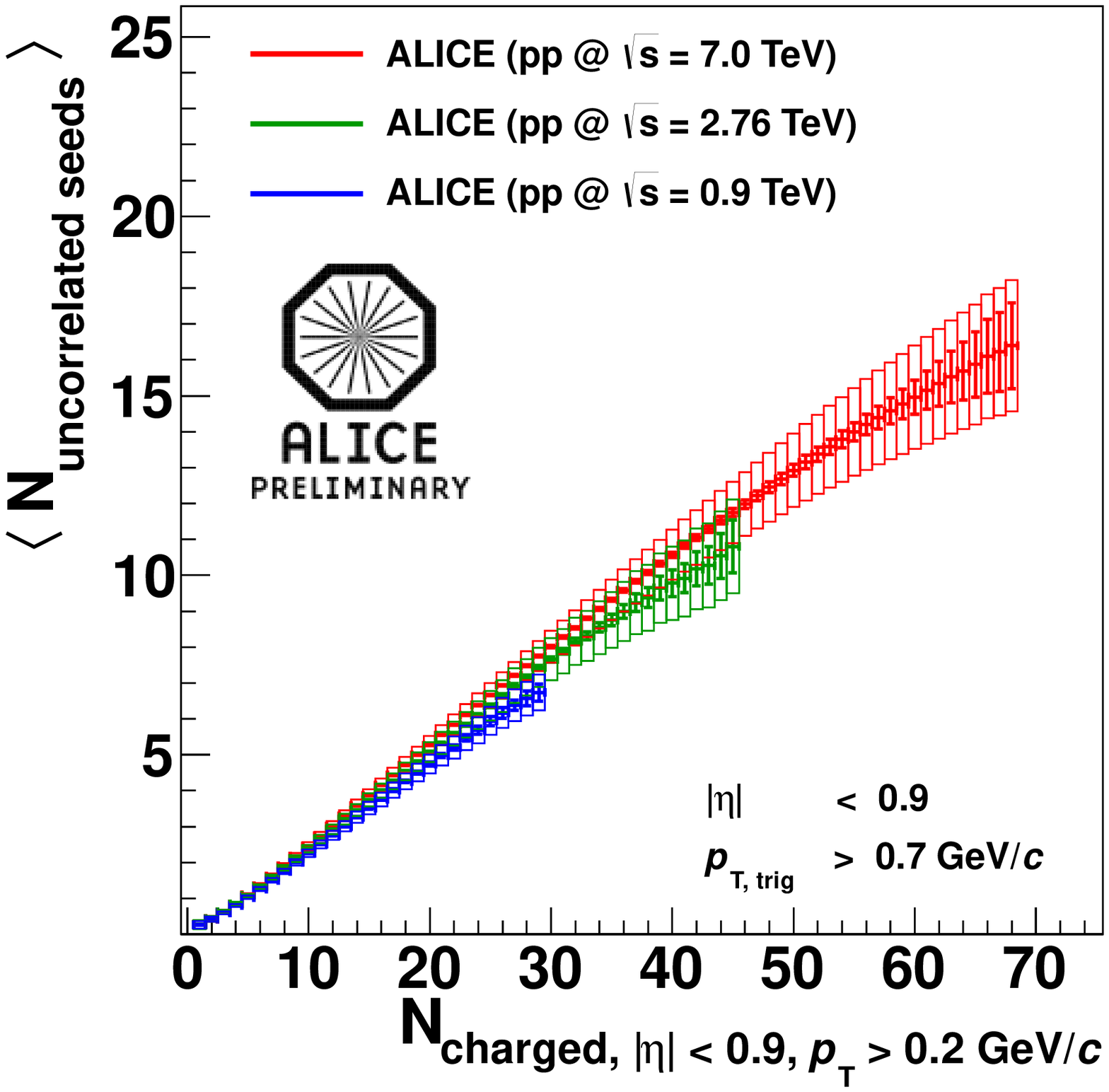,width=70mm}
\psfig{figure=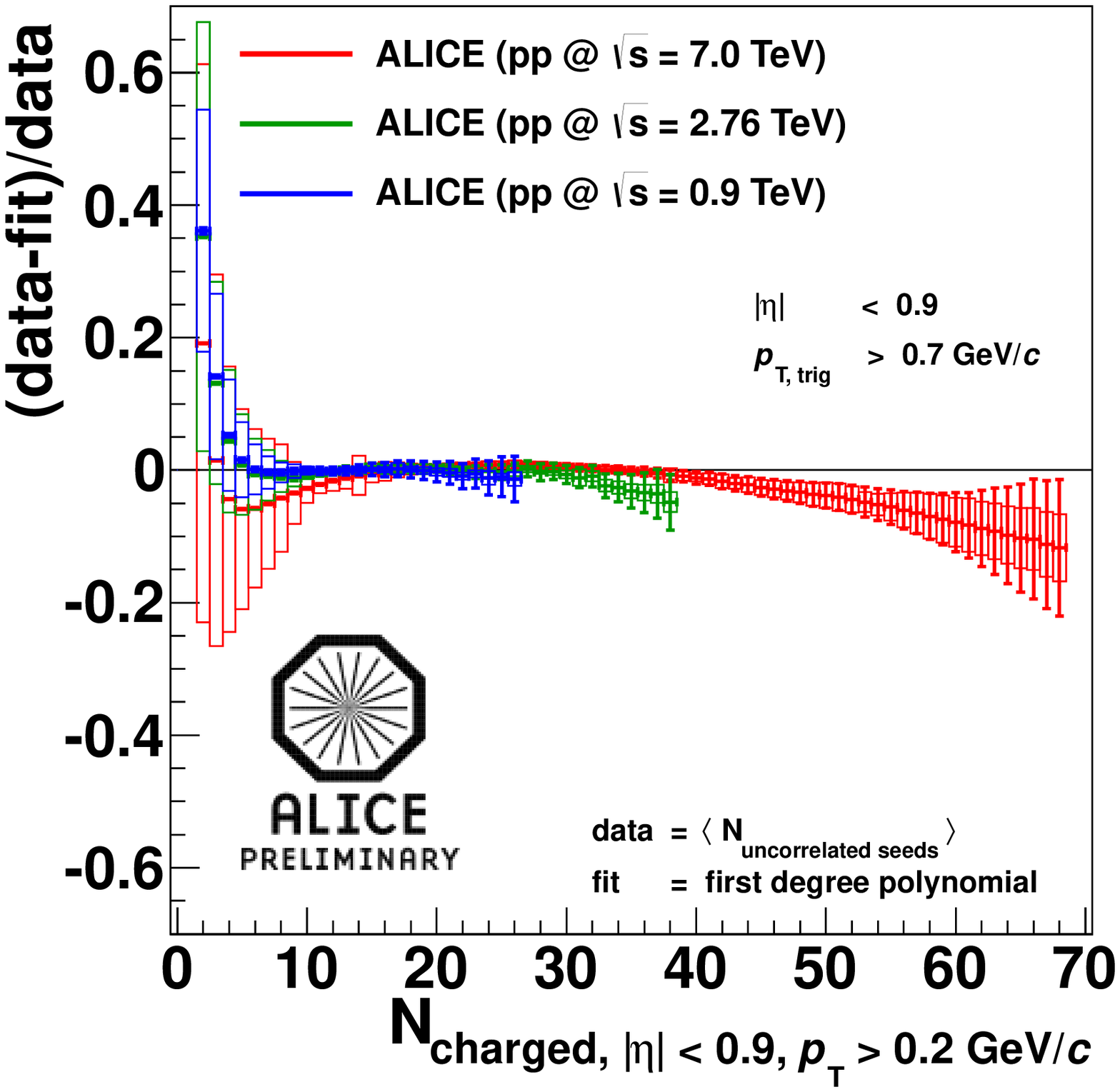,width=70mm}
\caption{Left panel: Average number of uncorrelated seeds as a function of the charged particle multiplicity. Right panel: Residuum between the number of uncorrelated seeds and linear fit functions.}
\label{fig:MPI}
\end{center}
\end{figure}

The analysis is also performed for pp collision data at $\sqrts=0.9$ and 2.76\,TeV (figures not shown). With decreasing collision energy, the agreement between the model predictions and the ALICE results increases. Also, the dependence of the correlation results on the transverse momentum threshold is studied (figures not shown). Here, we find that the model predictions over- or underestimate the near side and the away side yields of the ALICE results not only when using low transverse momentum thresholds, but over a large range of transverse momentum thresholds~($p_{\mathrm{T,(trig, assoc)}}>0.7$\,GeV/$c$ to $p_{\mathrm{T,(trig, assoc)}}>3.0$\,GeV/$c$). In general, Pythia Perugia-2011 gives the best description for all observables at the three studied collision energies, however, it does not fully agree with all observables.\\
In addition to the presented results, the dependence of the observables as a function of the center-of-mass energy is studied. While the combinatorial background in a fixed charged particle multiplicity bin does not show any center-of-mass energy dependence, the near side yield and the away side yield show an energy dependence (figures not shown). The energy dependence of all observables is roughly reproduced by all Pythia tunes. Only Phojet is in clear disagreement with the energy dependence of one ALICE result: Phojet does not show any energy dependence of the away side yield in the studied center-of-mass energy range. \\ 
Interpreted in the context of the Pythia model, the number of uncorrelated seeds gives information about the number of multiple parton interactions. In the left panel of Figure \ref{fig:MPI}, the average number of uncorrelated seeds as a function of the charged particle multiplicity is presented for the center-of-mass energies $\sqrts=0.9$, 2.76, and 7.0\,TeV. The right panel of Figure \ref{fig:MPI} shows the residuum between the data points and linear fit functions ((data-fit)/data). 
It can be observed that the charged particle multiplicity increases approximately linearly with the number of uncorrelated seeds. 
However, it deviates from the linear dependence at large charged particle multiplicities. Here, the rise of the number of uncorrelated seeds levels off. 
This observation is consistent with the assumption that there is a limit in the number of multiple parton interactions in pp collisions as predicted by Pythia.
\vspace{-0.3cm}

In summary, a new analysis method for the study of the jet fragmentation and the number of multiple parton interactions has been established. This analysis method has been successfully used to gain a better understanding of the particle production mechanisms in proton-proton collisions. Based on these results, theoretical models can be optimized at LHC energies.

\end{document}